\documentclass[manuscript]{aastex63}
\usepackage{amsmath}
\usepackage{graphicx}
\usepackage{slashed}
\shorttitle{{ \footnotesize Response tensor for a spin-dependent electron gas: dependence on the choice of spin operator}}
\shortauthors{Demcsak \& Melrose}

\begin{document}
\title{Response tensor for a spin-dependent electron gas: dependence on the choice of spin operator}
\email{victor.demcsak@sydney.edu.au}
\author{Victor M.~Demcsak}
\author{Donald B.~Melrose}
\affil{School of Physics, The University of Sydney, Sydney, NSW 2006, Australia}

\begin{abstract}
\noindent It is shown that the choice of spin-operator affects the form of the response tensor describing a spin-dependent electron gas. The covariant, spin-dependent response tensor for a magnetic dipole moment-polarized electron gas (statistical distribution of electrons and positrons) is evaluated. A simultaneous eigenfunction of both the magnetic-moment spin-operator and the Dirac Hamiltonian is constructed, from which explicit expressions for the magnetic-moment states and the corresponding vertex functions are derived. It is shown that a gas of electrons having a preferred magnetic-moment spin has a rotatory-type response that is gyrotropic. In contrast, when the helicity is chosen as the spin operator, the response of an electron gas with a preferred helicity spin has a rotatory response that is analogous to an optically active medium. The distinction between these spin operators does not appear in conventional treatments of spin-dependence in quantum plasmas.
\end{abstract}

\section{Introduction}
The response (to an electromagnetic disturbance) of quantum relativistic spin-dependent plasmas has been discussed in the literature using different treatments. There are three main approaches. The simplest approach is quantum fluid theory (Q\textit{f}T) \cite{haas2011quantum}, which is analogous to the fluid description of a classical plasma. Another approach is plasma kinetic theory (PKT) \cite{bonitz2016quantum}, which is based on the Wigner function \cite{wigner1997quantum}. Both Q\textit{f}T and PKT can be generalised to include spin and relativistic effects \cite{hurst2017phase, ekman2019relativistic}. Spin is included using the Schr\"odinger-Pauli theory, in which the spin operator is the vector with components given by the three Pauli matrices, $\sigma_x, \; \sigma_y,$ \textrm{and} $\sigma_z$. Relativistic effects are included by replacing Schr\"odinger's equation by Dirac's equation. 
\\
\\
Complications arise in the generalisation of both Q\textit{f}T and PKT. In particular how the spin is treated: the Dirac wavefunction is a column vector in the 4-dimensional Dirac spin space, with the four components (for a particle at rest) interpreted as an electron or a positron with spin up or spin down. 
A complication is that in general one cannot readily identify wavefunctions for a single electron or a single positron. A conventional approach is to make the Foldy-Wouthuysen (FW) transformation, which involves an infinite expansion. In practice this expansion needs to be truncated, and then the separation of electron and positron states is only approximate. The conventional approach to treating the spin is then to introduce Pauli-type matrices in both the $2\times2$ subspaces that describe either an electron or a positron. One can then interpret the spin by analogy with the Schr\"odinger-Pauli theory in both subspaces. This procedure defines a $4\times4$ spin operator that is an outer product of two $2\times2$ Pauli matrices. However, this spin operator does not commute with the Dirac Hamiltonian $(\hat{H})$, so that the wavefunctions do not retain their (postulated) form as time evolves. 
\\
\\
It is a necessary condition that any relativistically acceptable spin operator must commute with the Dirac Hamiltonian. More generally, the Dirac spin operator, $S^{\mu \nu} = 1/2(\gamma^{\mu} \gamma^{\nu} - \gamma^{\nu} \gamma^{\mu} )$, which is both a $4\times4$ matrix and a second rank 4-tensor, does not commute with $\hat{H}$. Two examples of relativistically acceptable spin operators are: the magnetic dipole moment operator $(\hat{\mu})$, and the helicity operator $(\hat{\sigma})$ \cite{sokolov1966synchrotron, sokolov1986radiation, fradkin1961electron}.
\\
\\
In this paper we study relativistic quantum plasmas using the framework of quantum plasmadynamics (QPD) \cite{melrose1996quantum,melrose2008quantum,melrose2012quantum}. In QPD, the response to an electromagnetic disturbance for an  electron gas can be written in a manifestly covariant form as a generalization of the vacuum polarization with the electron propagator in vacuo replaced by a statistically averaged electron propagator \cite{melrose2008quantum}. The statistical average involves occupation numbers $n^{\epsilon}_{s}(\vec{p})$ describing electrons ($\epsilon = + 1$), and positrons ($\epsilon = - 1$), with $s=\pm1$ the eigenvalues of a spin operator that commutes with the Dirac Hamiltonian.  Natural units ($c=\hbar=1$) are used throughout this paper.
\\
\\
We derive the linear response 4-tensor $\Pi_{\textrm{sd}}^{\mu\nu}$, corresponding to a  spin-dependent electron gas with the spin operator identified as the $z$-component of the magnetic-dipole-moment operator. One motivation is to compare the response for a magnetic-moment polarized electron gas with that of a spin-polarized electron gas in which the spin operator is the helicity \cite{melrose2003optical}. In the helicity-dependent case, $\Pi_{\textrm{sd}}^{\mu\nu}$ implies a rotatory response analogous to a solution of dextrose, which can be described as optically active. This is plausible in that an electron (or positron) with positive helicity has a handedness analogous to the handedness of a molecule of dextrose. The question that we aim to answer is how the spin-dependent part of the response depends on the choice of spin operator. This question does not arise in theories, such as those mentioned above, in which the spin operator is Pauli-like and ``spin-dependence'' does not involving distinguishing between different spin operators (that commute with the Dirac Hamiltonian).
\\
\\
An isotropic magnetic-dipole-moment-dependent electron gas is qualitatively different from an isotropic helicity-dependent electron gas in that the latter has a preferred axis, here the $z$-axis, whereas there is no preferred axis in the helicity-dependent case. For example, a magnetic-dipole-moment-dependent electron gas could be constructed by separating spin up from spin down in an external magnetic field along the $z$-axis and collecting only those particles with spin up. We are interested in how the existence of this direction affects the spin-dependent part of the response tensor. 
\\
\\
In Section \ref{213367} we identify the response tensor for an arbitrary electron gas that is unmagnetized. The response tensor of a spin-dependent electron gas can be separated into a spin-dependent part and a spin-independent part. The spin-independent part can be evaluated without the introduction of a spin-operator, specifically by taking the trace over a product of Dirac matrices. In Section \ref{2133} we identify a simultaneous eigenfunction of the Dirac Hamiltonian, and the magnetic-moment operator. The corresponding vertex function is derived, and is used to evaluate the spin-dependent part of the response tensor describing a magnetic-moment polarized electron gas. It is shown that this response tensor is of the rotatory \cite{melrose2005electromagnetic} form discussed in Section \ref{213367}.  The difference between the responses of the helicity and magnetic-moment-dependent cases is discussed in Section \ref{diff378}.

\section{Response of an unmagnetized electron gas} \label{213367}
Quantum Plasmadynamics (QPD) \cite{melrose2008quantum,melrose2012quantum} introduces a covariant description for the (linear) response tensor. The disturbance is described by the 4-potential $A^{\mu}(k)$, and the induced 4-current $J^{\mu}(k)$ describes the response, where the argument $k$ is the wave 4-vector $k^{\mu}=(\omega, \vec{k})$. The linear response tensor $\Pi^{\mu\nu}(k)$ relates these two quantities $J(k)^{\mu}=\Pi^{\mu\nu}(k)A_{\nu}(k)$. Charge continuity and gauge invariance impose the following conditions respectively:
\begin{equation}
k_{\mu} \Pi^{\mu \nu}(k)=0, \;\; k_{\nu}\Pi^{\mu \nu}(k)=0.
\label{gaugecharge}
\end{equation}
The higher order nonlinear response tensors follow by considering the electron loops having $n+1$ vertices. The response tensor gives a complete description of the electromagnetic properties of a given medium. The mathematical constraints on the response tensor indicate the physical requirements of the medium.
\subsection{The covariant description of isotropic media}
The linear response $\Pi^{\mu \nu}(k)$ can be written as a sum of components, where each component is the product of an invariant function with a second rank tensor. The possible second rank tensors for isotropic media are the 4-vector $k^{\mu}$, 4-velocity $v^{\mu}$, the metric tensor $g^{\mu \nu}$ (in this paper the signature $-1$ is used), and the completely asymmetric tensor $\epsilon^{\mu \nu \rho \sigma}$. The allowed products of the invariant function with an appropriate second rank tensor are constrained by the reality condition $\Pi^{\mu \nu}(k)=[\Pi^{\mu \nu}(-k)]^{*}$, charge continuity and gauge invariance (Eq. \ref{gaugecharge}), and the Onsager relations. In particular the Onsager relations require that the time reversed non-dissipative part of the response needs to be an even function $\bar{\Pi}^{H_{\mu \nu}} (\bar{k})|_{\bar{F}_{0}} = \Pi^{H_{\mu \nu}}(k)|_{F_{0}}$, where $H$ indicates a hermitian quantity, the bar indicates a time reversal, and $\bar{F}_{0}$ indicates a time reversed static field. Similarly the dissipative part (involving an anti-hermitian part) must be an odd function. 
\\
\\
For an isotropic medium, one can only construct three independent tensors satisfying the above conditions, and therefore at most three invariants are required to completely describe the response of an arbitrary isotropic medium. The general form of the response tensor for an isotropic medium takes the following form \cite{d1989electromagnetic}:
\begin{eqnarray}
\Pi^{\mu \nu}(k) =&& \Pi^{L}(k) L^{\mu \nu}(k, u) + \Pi^{T}(k) T^{\mu \nu}(k, u) \nonumber \\
&&+  \Pi^{R}(k) R^{\mu \nu}(k, u),
\label{generalformresponse}
\end{eqnarray}
where $u^\mu$ is the 4-velocity of the rest frame of the plasma: $L^{\mu \nu}, T^{\mu \nu}, R^{\mu \nu}$ respectively define the longitudinal, transverse and rotatory parts, and $\Pi^{L}, \Pi^{T}, \Pi^{R}$ are their corresponding response functions. The tensors in Eq. \ref{generalformresponse} can be thought of as the projection-like operators for $\Pi^{L}, \Pi^{T}, \textrm{and}\; \Pi^{R}$.
\\
\\
The longitudinal tensor may be written as the outer product of two 4-vectors: $L^{\mu \nu}(k, v)=-L^{\mu}(k, v)L^{\nu}(k, v)$, where this is normalised in such a way that in the rest frame $u = (1, \vec{0}\;)$ one has $L(k, u) = (|\vec{k}|/\omega, \vec{k}/|\vec{k}|)$ and,
\begin{equation}
L^{\mu}(k,u) = \frac{ku k^{\mu} - k^{2} u^{\mu}}{ku \sqrt{(ku)^{2} - k^{2}}}.
\end{equation}
The transverse tensor is normalized such that it takes the form of the unit transverse second rank tensor in the rest frame:
\begin{equation}
T^{\mu \nu}(k,u) = \frac{(ku)^{2}}{k^{2}} L^{\mu}(k, u) L^{\nu}(k, u) + g^{\mu \nu} - \frac{k^{\mu} k^{\nu}}{k^{2}}.
\end{equation}

\noindent The rotatory tensor is
\begin{equation}
R^{\mu \nu}(k,v) = i \epsilon ^{\mu \nu \rho \sigma} L_{\rho}(k) v_{\sigma}.
\end{equation}

\noindent The invariants can be constructed from the linear response for an isotropic plasma, and they take the following forms:
\begin{align}
\Pi^{L}(k) &= \frac{(k v)^{4}}{k^{4}} L_{\mu \nu}(k, v) \Pi^{\mu \nu}(k), \\
\Pi^{T}(k) &= \frac{1}{2} T_{\mu \nu}(k, v) \Pi^{\mu \nu}(k),\\
\Pi^{R}(k) &= -\frac{1}{2} R_{\mu \nu} (k, u) \Pi^{\mu \nu}(k).
\end{align}
\subsection{Response of an arbitrary electron gas}
The polarization of the vacuum field can be calculated using the electron propagator in vacuo. In an analogous way, the general response tensor can be computed using QED by replacing the vacuum electron propagator with a statistically averaged propagator describing the electron gas. This average involves occupation numbers $n^{\epsilon}_{s}(\vec{p})$ describing electrons ($\epsilon = + 1$), and positrons ($\epsilon = - 1$), where $s=\pm1$ is the spin eigenvalue. The general response of an arbitrary electron gas, derived in \cite{melrose1983quantum}, has the form:
\begin{eqnarray}
\Pi^{\mu \nu} (k) =&& -e^{2} \sum_{\epsilon, \epsilon', s, s'} \int \frac{d^{3}\vec{p}}{(2\pi)^{3}} \int \frac{d^{3}\vec{p}\;'}{(2\pi)^{3}}(2\pi)^{3} \nonumber \\  &&\times \delta^{3} (\epsilon' \vec{p}\;' - \epsilon \vec{p} + \vec{k} ) \frac{\epsilon n_{s}^{\epsilon}(\vec{p}\;) - \epsilon' n_{s'}^{\epsilon'}(\vec{p}\;')}{\omega - \epsilon \varepsilon + \epsilon' \varepsilon' + i0}  \nonumber \\
&& \times [\Gamma _{s'\; s}^{\epsilon' \; \epsilon}(\vec{p\hspace{0.5mm}}\hspace{0.5mm}', \vec{p\hspace{0.5mm}}, \vec{k\hspace{0.5mm}} )]^{\mu}
[\Gamma _{s'\; s}^{\epsilon' \; \epsilon}(\vec{p\hspace{0.5mm}}\hspace{0.5mm}', \vec{p\hspace{0.5mm}}, \vec{k\hspace{0.5mm}} )]^{*\nu},
\label{4378}
\end{eqnarray}
with $\varepsilon = \sqrt{\vec{p}\;^{2} + m^{2}}$, and $\varepsilon' = \sqrt{\vec{p}^{\; \prime \hspace{0.5mm} 2} + m^{2}}$. The vertex functions used to calculate the linear response tensor are defined as:
\begin{equation}
[\Gamma _{s'\; s}^{\epsilon' \; \epsilon}(\vec{p\hspace{0.5mm}}\hspace{0.5mm}', \vec{p\hspace{0.5mm}}, \vec{k\hspace{0.5mm}} )]^{\mu} = \frac{\bar{u}_{s'}^{\epsilon'}(\epsilon' \vec{p}\;') \gamma^{\mu} u_{s}^{\epsilon}(\epsilon \vec{p}\;)}{\sqrt{4\varepsilon\varepsilon'}},
\label{vertex11}
\end{equation}
where $\gamma^{\mu}$ are the Dirac matrices, and $u_{s}^{\epsilon}$, $\bar{u}_{s'}^{\epsilon'}$ are the electron and positron eigenfunctions respectively.

\subsection{Spin-dependent and spin-independent contributions}
The occupation number can be separated into spin-averaged and spin-specific components. This is achieved by explicitly identifying the choice of spin eigenvalue for a given $\epsilon$ in the following way:
\begin{equation}
n^{\epsilon} (\vec{p}) = \frac{1}{2} [n_{+}^{\epsilon} (\vec{p}\;) + n_{-}^{\epsilon} (\vec{p}\;)],
\end{equation}
\begin{equation}
\Delta n^{\epsilon} (\vec{p}) = \frac{1}{2} [n_{+}^{\epsilon} (\vec{p}\;) - n_{-}^{\epsilon} (\vec{p}\;)].
\end{equation}

\noindent It follows that the linear response tensor in Eq. \ref{4378} separates into a spin-dependent part $\Pi^{\mu \nu}_{\textrm{sd}}(k)$, and a spin-independent part $\Pi^{\mu \nu}_{\textrm{in}}(k)$, such that $\Pi^{\mu \nu}(k)= \Pi^{\mu \nu}_{\textrm{sd}}(k) +\Pi^{\mu \nu}_{\textrm{in}}(k)$. To see this notice that we have in Eq. \ref{4378}:
\begin{eqnarray}
&&\sum_{s, s'} n_{s}^{\epsilon} (\vec{p}\;) [\Gamma _{s'\; s}^{\epsilon' \; \epsilon}(\vec{p\hspace{0.5mm}}\hspace{0.5mm}', \vec{p\hspace{0.5mm}}, \vec{k\hspace{0.5mm}} )]^{\mu} [\Gamma _{s'\; s}^{\epsilon' \; \epsilon}(\vec{p\hspace{0.5mm}}\hspace{0.5mm}', \vec{p\hspace{0.5mm}}, \vec{k\hspace{0.5mm}} )]^{*\nu} \nonumber \\
&&= n^{\epsilon}(\vec{p}\;)\mathcal{A} + \Delta n^{\epsilon} (\vec{p}\;) \mathcal{B},
\end{eqnarray}

\begin{eqnarray}
&&\sum_{s, s'}  n_{s'}^{\epsilon'} (\vec{p}\;') [\Gamma _{s'\; s}^{\epsilon' \; \epsilon}(\vec{p\hspace{0.5mm}}\hspace{0.5mm}', \vec{p\hspace{0.5mm}}, \vec{k\hspace{0.5mm}} )]^{\mu} [\Gamma _{s'\; s}^{\epsilon' \; \epsilon}(\vec{p\hspace{0.5mm}}\hspace{0.5mm}', \vec{p\hspace{0.5mm}}, \vec{k\hspace{0.5mm}} )]^{*\nu} \nonumber \\
&&= n^{\epsilon'}(\vec{p}\;')\mathcal{A} + \Delta n^{\epsilon'} (\vec{p}\;') \mathcal{C},
\end{eqnarray}
with
\begin{align}
\mathcal{A} &= \sum_{s, s'} \;[\Gamma _{s'\; s}^{\epsilon' \; \epsilon}(\vec{p\hspace{0.5mm}}\hspace{0.5mm}', \vec{p\hspace{0.5mm}}, \vec{k\hspace{0.5mm}} )]^{\mu}
[\Gamma _{s'\; s}^{\epsilon' \; \epsilon}(\vec{p\hspace{0.5mm}}\hspace{0.5mm}', \vec{p\hspace{0.5mm}}, \vec{k\hspace{0.5mm}} )]^{*\nu},\\
\mathcal{B} &= \sum_{s, s'} \; s \;[\Gamma _{s'\; s}^{\epsilon' \; \epsilon}(\vec{p\hspace{0.5mm}}\hspace{0.5mm}', \vec{p\hspace{0.5mm}}, \vec{k\hspace{0.5mm}} )]^{\mu}
[\Gamma _{s'\; s}^{\epsilon' \; \epsilon}(\vec{p\hspace{0.5mm}}\hspace{0.5mm}', \vec{p\hspace{0.5mm}}, \vec{k\hspace{0.5mm}} )]^{*\nu},\label{whereB}\\
\mathcal{C} &= \sum_{s, s'} \; s' \;[\Gamma _{s'\; s}^{\epsilon' \; \epsilon}(\vec{p\hspace{0.5mm}}\hspace{0.5mm}', \vec{p\hspace{0.5mm}}, \vec{k\hspace{0.5mm}} )]^{\mu}
[\Gamma _{s'\; s}^{\epsilon' \; \epsilon}(\vec{p\hspace{0.5mm}}\hspace{0.5mm}', \vec{p\hspace{0.5mm}}, \vec{k\hspace{0.5mm}} )]^{*\nu}. \label{whereC}
\end{align}
\\
The spin-independent part of the response tensor is \cite{hayes1984dispersion}:
\begin{eqnarray}
\Pi_{\textrm{in}}^{\mu \nu}(k) &&=2e^{2} \int \frac{d^{3}\vec{p}}{(2\pi)^{3}} \sum_{\epsilon} \frac{n^{\epsilon}(\vec{p}\;)}{\varepsilon} \nonumber \\
&& \times \left [\frac{F^{\mu \nu}(\epsilon p, \epsilon p -k)}{k^{2} - 2\epsilon pk}+ \frac{F^{\mu \nu}(\epsilon p +k, \epsilon p )}{k^{2} + 2\epsilon pk}\right], 
\end{eqnarray}
where $F^{\mu \nu}(\epsilon \tilde{p}, \epsilon' \tilde{p}') = \epsilon \varepsilon \epsilon' \varepsilon' \mathcal{A}$, and $\tilde{p} = [\varepsilon, \vec{p}]$, $\tilde{p}\;' = [\varepsilon ', \vec{p}\; ']$.
\\
\\
There are two different options to evaluate $\Pi^{\mu \nu}_{\textrm{in}}(k)$. One way is to introduce a spin operator and directly compute it using a specified vertex function. In particular $\mathcal{A}$ is the same regardless of the choice of spin operator (provided that it commutes with the Dirac Hamiltonian, and is well defined under a Lorentz transform). Alternatively one can evaluate the spin-independent response without explicitly specifying any spin operator by evaluating the following trace written using the Feynman slash notation:
\begin{equation}
F^{\mu \nu} (K, K') = \frac{1}{4} \textrm{Tr} [\gamma^{\mu}(\slashed{K}+m) \gamma^{\nu}(\slashed{K}' +m)].
\end{equation}

\noindent The spin-dependent part of the response tensor is the sum of two contributions:
\begin{eqnarray}
\Pi_{\textrm{sd}}^{\mu\nu}(k) &&= -e^{2} \sum_{\epsilon, \epsilon'} \int \frac{d^{3} \vec{p}}{(2\pi)^{3}} \frac{\epsilon \Delta n^{\epsilon}(\vec{p})}{\omega - \epsilon\varepsilon + \epsilon' \varepsilon' + i0} \; \mathcal{B} \nonumber \\
&& +e^{2} \sum_{\epsilon, \epsilon'} \int \frac{d^{3} \vec{p}\hspace{0.5mm}'}{(2\pi)^{3}} \frac{\epsilon' \Delta n^{\epsilon'}(\vec{p}\hspace{0.5mm}')}{\omega - \epsilon\varepsilon + \epsilon' \varepsilon' + i0} \;\mathcal{C}.
\label{sdtensor}
\end{eqnarray}

\section{Response of a magnetic-moment-dependent electron gas} \label{2133}
In this section we evaluate Eq. \ref{whereB} and Eq. \ref{whereC} which will contribute to the first and second parts of the spin-dependent linear response tensor $\Pi^{\mu \nu}_{\textrm{sd}}(k)$ respectively. Both $\mathcal{B}$ and $\mathcal{C}$ are explicitly dependent on the choice of spin operator. The nonzero components of $\mathcal{B}$ and $\mathcal{C}$ give the magnetic moment-dependent part of the response tensor. 

\subsection{Magnetic moment eigenfunction}
The temporal evolution of the magnetic moment operator $\hat{\vec{\mu}}$ is governed by ${d \hat{\vec{\mu}}}/{dt} = i[\hat{H}, \hat{\vec{\mu}}]$, where $\hat{H}$ is the Dirac Hamiltonian. The inclusion of the electromagnetic field recovers $ {d \hat{\vec{\mu}}}/{dt} = e\gamma^{0} \vec{\sigma} \times \vec{B} - ie\vec{\gamma} \times \vec{E}$, where $\vec{\sigma}$ are the Pauli matrices in the $4\times 4$ Dirac spin space. Therefore $\hat{\mu_{z}}$ is a constant of the motion if a magnetostatic field is present along the $z$-axis. The $z$-component of the magnetic moment operator takes the following form:
\begin{equation}
\hat{\mu}_{z}=
\left(
\begin{array}{cccc}
\epsilon m & 0 & 0 & p_{\perp}e^{-i\phi}\\
0 & -\epsilon m & -p_{\perp} e^{i\phi} & 0\\
0 & -p_{\perp} e^{-i\phi} & \epsilon m & 0\\
p_{\perp} e^{i\phi} & 0 & 0 & -\epsilon m\\
\end{array}
\right),
\label{mmoment0}
\end{equation}

\noindent and its eigenvalues are $s\lambda$, where $s=\pm1$ and $\lambda = \sqrt{m^{2} + p_{\perp}^{2}}$. The momentum is expressed using cylindrical polar coordinates $\vec{p} = (p_{\perp} \cos{\phi}, p_{\perp} \sin{\phi}, p_{z})$.
\\
\\
The spin operator $\hat{\mu}_{z}$ is a relativistically acceptable spin operator \cite{sokolov1966synchrotron, sokolov1986radiation, fradkin1961electron}, since it commutes with the Dirac Hamiltonian which ensures the eigenvalues of $\hat{\mu}_{z}$ are constants of the motion. The polarization states correspond to the eigenfunctions of the component of the magnetic-moment operator $\hat{\vec{\mu}}$ along the direction of the magnetic field $\vec{b}$ used to separate the up and down spin states in creating the plasma.
\\
\\
 We assume plane-wave solutions of the form $\exp{[-i\epsilon (\varepsilon t - \vec{p} \cdot \vec{x})]}$, where $\hat{\vec{p}} = -i\partial / \partial\vec{x}$ in the coordinate representation. The sign of $p_{z}$ is written $\mathbb{P} = p_{z}/|p_{z}|$.  One can then rewrite the electron/positron wavefunction more conveniently by using the identities:
\begin{align}
p_{z} &= \mathbb{P} \sqrt{\varepsilon + s\epsilon \lambda} \sqrt{\varepsilon - s\epsilon \lambda} \label{useinappendix1},\\
p_{\perp} &= \sqrt{\lambda+sm}\sqrt{\lambda-sm}.
\end{align} 

\noindent A specific choice of simultaneous eigenfunctions of both $\hat{\mu}_{z}$, and the Dirac Hamiltonian is:
\begin{equation}
u_{s}^{\epsilon}(\epsilon \vec{p} \hspace{0.7mm})=
\frac{1}{\sqrt{4 \lambda  \varepsilon}}
\left(
\begin{array}{c}
\sqrt{\varepsilon + \epsilon s \lambda} \; \sqrt{\lambda+ sm}\; e^{- i\phi/2} \\
- \mathbb{P} s \epsilon \; \sqrt{\varepsilon - \epsilon s p} \; \sqrt{\lambda- sm}\; e^{ i\phi/2} \\
\mathbb{P} \; \sqrt{\varepsilon - \epsilon s \lambda} \; \sqrt{\lambda+ sm}\; e^{- i\phi/2}\\
s \epsilon \; \sqrt{\varepsilon + \epsilon s \lambda} \; \sqrt{\lambda- sm}\; e^{ i\phi/2}\\
\end{array}
\right).
\label{mmoment1}
\end{equation}

\noindent The vertex function (Eq. \ref{vertex11}) for the magnetic moment eigenfunction (Eq. \ref{mmoment1}) is given by 
\begin{widetext}
\begin{equation}
[\Gamma_{s, s'}^{\epsilon, \epsilon'} (\vec{p}\;', \vec{p}\;)]^{\mu}=
\frac{1}{\sqrt{16 \lambda\lambda' \varepsilon \varepsilon'}}\\
\left(
\begin{array}{cccc}
[\alpha _{+} ' \alpha_{+} + \mathbb{P}^{\hspace{0.3mm }\prime} \hspace{0.5mm} \mathbb{P} \alpha_{-}' \alpha_{-}] [\beta_{+}' \beta_{+} e^{-i(\phi - \phi')/2} + \Sigma \beta_{-}' \beta_{-} e^{i(\phi - \phi')/2}]  \\

[\alpha _{+} ' \alpha_{+} - \mathbb{P}^{\hspace{0.3mm }\prime} \hspace{0.5mm} \mathbb{P} \alpha_{-}' \alpha_{-}][\epsilon s \beta_{+}' \beta_{-} e^{i(\phi + \phi')/2} + \epsilon' s' \beta_{-}' \beta_{+} e^{-i(\phi + \phi')/2}] \\

-i[\alpha _{+} ' \alpha_{+} - \mathbb{P}^{\hspace{0.3mm }\prime} \hspace{0.5mm} \mathbb{P} \alpha_{-}' \alpha_{-}] [\epsilon s \beta_{+}' \beta_{-} e^{i(\phi + \phi')/2} - \epsilon' s' \beta_{-}' \beta_{+} e^{-i(\phi+\phi')/2}]\\

\mathbb{P}[\alpha _{+} ' \alpha_{-} + \mathbb{P}^{\hspace{0.3mm }\prime} \hspace{0.5mm} \mathbb{P} \alpha_{-}' \alpha_{+}] [\beta_{+}' \beta_{+} e^{-i(\phi - \phi')/2} + \Sigma \beta_{-}' \beta_{-} e^{i(\phi - \phi')/2}]\\
\end{array}
\right),
\label{vertexmagmoment}
\end{equation}
where
$a_{\pm} = \sqrt{\varepsilon \pm s\epsilon \lambda} $,
$ b_{\pm} = \sqrt{\lambda \pm sm}\; e^{\mp i\phi/2}$,
$\Sigma = s s' \epsilon \epsilon'$,
$a'_{\pm} = \sqrt{\varepsilon' \pm s'\epsilon' \lambda'} $, and
$ b'_{\pm} = \sqrt{\lambda' \pm s'm}\; e^{\mp i\phi'/2}$.


\subsection{The response tensor for magnetic-moment eigenstates}

Direct evaluation of Eq. \ref{sdtensor} recovers the magnetic-moment dependent part of the response tensor.

 \begin{equation}
\Pi_{\textrm{sd}}^{\mu \nu}=
i m e^{2} k^{2}  \sum_{\epsilon} \int \frac{d^{3} \vec{p}}{(2\pi)^{3}}\frac{\epsilon \Delta n^{\epsilon}(\vec{p})}{(pk)^{2} - ({k^{2}}/{2})^{2}} \;
b^{\mu \nu}(k,p),
\nonumber 
\end{equation}
 
 \noindent where 
 
\begin{eqnarray}
b^{\mu \nu} (k,p)=
\left(
\begin{array}{cccc}
0 & k_{y}/p & -k_{x}/p &0 \\
-k_{y}/p & 0 & (-\epsilon \omega + p_{z} k_{z})/\varepsilon p & -p_{z}k_{y}/\varepsilon p \\
k_{x}/p & (\varepsilon \omega - p_{z}k_{z})/\varepsilon p & 0 & p_{z}k_{x}/\varepsilon p \\
0 & p_{z}k_{y}/\varepsilon p & -p_{z}k_{x}/\varepsilon p &0 \\
\end{array}
\right),
\label{spinmagresponse2}
\end{eqnarray}

 \noindent with $pk = \omega \varepsilon - \vec{p} \cdot \vec{k}$, and $k^{2} = \omega^{2} - |\vec{k}\hspace{0.5mm}|^{2}$. A sample derivation is given in the Appendix.
 \end{widetext}
{\subsection{Rotatory Response} \label{diff378}}
The spin-dependent response tensor for a magnetic-dipole-moment polarized electron gas is antisymmetric and of the rotatory form. A magnetic-dipole-moment polarized electron gas is a gyrotropically active plasma, analogous to a cold magnetised electron gas. Before considering the general case, it is instructive to investigate a particular case. Select $\vec{k}$ along a particular axis in the rest frame of the electron gas. Choosing along the \textit{z}-axis $\vec{k}=(0, 0, |\vec{k}\hspace{0.5mm}|)$ implies that $\vec{p} \cdot \vec{k} = p_{z}|\vec{k}\hspace{0.5mm}|$. It follows that the integrals over terms that depend on $p_{x}$ or $p_{y}$ in Eq. \ref{spinmagresponse2} give zero, thus the only non-zero terms in Eq. \ref{spinmagresponse2} are $b^{12}=-b^{21}$. Therefore the only non-zero component of the rotatory tensor when $\vec{k}$ is along the $z$-axis in the rest frame is $b^{12}$. 
\\
\\
The general case can be investigated by noting that the spin-dependent part of the response tensor for both the helicity and magnetic-moment operators may be written in the form
$b^{\mu\nu}=\varepsilon^{\mu\nu\rho\sigma}k_\rho g_\sigma$.
For the helicity case one has
\begin{equation}
g_\sigma\propto p_\sigma+\frac{m^2}{\varepsilon}u_\sigma,
\end{equation}
where $u^\mu$ is the 4-velocity of the plasma frame. In the rest frame
$u^\mu=(1,{\bf 0})$. By inspection of Eq. \ref{spinmagresponse2}, the response of a magnetic-moment polarised electron gas can be written in the form
\begin{equation}
g_\sigma\propto b_\sigma+\frac{p_zk_z}{\varepsilon}u_\sigma,
\end{equation}
with $b^{\mu}=(0,{\bf b})$ in terms of the unit vector ${\bf b}$ along the direction of the magnetic moment. An isotropic plasma has no preferred direction for ${\bf p}$. Upon averaging over ${\bf p}$, one has $\langle p_\sigma\rangle=\varepsilon u_\sigma$ and $\langle p_z \rangle=0$. The two cases give $g_\sigma\propto u_\sigma$ and  $g_\sigma\propto b_\sigma$, respectively. Both correspond to rotatory responses, with the handedness determined relative to ${\bf k}$ and to ${\bf b}$, respectively. 
\\
\\
One may describe the difference between the responses of the helicity and magnetic-moment cases as optically active, and gyrotropic respectively. In an isotropic plasma with helicity-polarized electrons there is no specific direction defined, and one can select ${\bf k}$ in any direction without losing generality. In contrast, the magnetic-moment operator in the present paper is ${\hat{\mu}_z}$, which involves a specific choice of the $z$-direction. 
\\
\\
\section{Discussion and Conclusions}

The result derived in this paper shows that the choice of spin operator affects the form of the spin-dependent part of the response of a spin-dependent electron gas. As already shown \cite{melrose2003optical}, if the spin operator is chosen to be the helicity operator the response of a spin-dependent isotropic electron gas has a rotatory component as in an optically active medium, such as a solution of dextrose. The natural modes of such a medium are circularly polarized. In contrast, we find that when the spin operator is chosen to be the magnetic-moment operator, the spin-dependent part of the response of an isotropic (in momentum) electron gas is gyrotropic, corresponding to a rotatory part along the axis that defines spin up and spin down. The natural modes of such a medium are elliptically polarized in general, reducing to circular and linear polarizations for propagation parallel and perpendicular to the axis that defines spin up and spin down.
\\
\\
Any spin-dependence is intrinsically quantum mechanical, and the spin-dependent part of the response is of order $\hbar$ smaller than the spin-independent part of the response. Consequently, it would be difficult to measure the effects, described here, that differentiate between the responses of different spin-polarized electron gases. The main implication of our result is that the concept of a ``spin-dependent'' electron gas is ill-defined in the absence of a clearly defined choice of an acceptable spin operator, which commutes with the Dirac Hamiltonian. Conventional treatments of spin-dependence in quantum plasmas make no distinction between these spin operators and involve a Pauli-like spin operator that does not commute with the Dirac Hamiltonian. Such treatments cannot identify differences between acceptable spin operators in relativistic quantum theory.

\section{Acknowledgements}
We thank Jeanette Weise for helpful advice and Michael Wheatland for helpful comments on the manuscript. VD is supported by the Australian Research Training Program.


\section{Appendix: Derivation of $b^{12}$}
The derivation of the 12-component of Eq. \ref{spinmagresponse2} proceeds as follows. We start with taking the product of the vertex functions corresponding to the tensor entry to be computed.
\begin{eqnarray*}
&&[\Gamma]^{1}[\Gamma]^{*2}=\frac{i}{16 \varepsilon p \varepsilon' p'} \times 
 \left[
2\varepsilon' \varepsilon + 2 s' \epsilon' p' s \epsilon p - 2 p_{z}' p_{z}
\right] \\
&& \times \left[ 2s' mp - 2p' sm - \left(e^{i(\phi' + \phi)} - e^{-i(\phi' + \phi)}\right)ss' \epsilon \epsilon' p_{\perp} p_{\perp}' \right],
\end{eqnarray*}
where
\begin{eqnarray}
\noindent \sum_{s, s'} s [\Gamma]^{1}[\Gamma]^{*2} &= \frac{im}{2\varepsilon \varepsilon' p}\left(-\varepsilon' \varepsilon + p^{2} \epsilon \epsilon' + p_{z}' p_{z} \right)
\end{eqnarray}
and
\begin{eqnarray}
\sum_{s, s'} s' [\Gamma]^{1}[\Gamma]^{*2} &= \frac{im}{\varepsilon \varepsilon' p'} (\varepsilon' \varepsilon - p'^{\hspace{0.5mm} 2} \epsilon \epsilon' - p_{z}' p_{z}).
\end{eqnarray}

The spin-dependent part of the response tensor is the sum of the two terms in Eq. \ref{sdtensor}

\begin{eqnarray}
\Pi^{12}_{\textrm{sd}} = -e^{2} \sum_{\epsilon, \epsilon'} \int && \frac{d^{3} \vec{p}}{(2\pi)^{3}} \frac{\epsilon \Delta n^{\epsilon}(\vec{p})}{\omega - \epsilon\varepsilon + \epsilon' \varepsilon' } \\ 
&& \times \frac{im}{2\varepsilon \varepsilon' p}\left(-\varepsilon' \varepsilon + p^{2} \epsilon \epsilon' + p_{z}' p_{z} \right) \nonumber\\
 +e^{2} \sum_{\epsilon, \epsilon'} \int && \frac{d^{3} \vec{p}\hspace{0.5mm}'}{(2\pi)^{3}} \frac{\epsilon' \Delta n^{\epsilon'}(\vec{p}\hspace{0.5mm}')}{\omega - \epsilon\varepsilon + \epsilon' \varepsilon' } \nonumber \\
 &&\times \frac{im}{\varepsilon \varepsilon' p'} (\varepsilon' \varepsilon - p'^{\hspace{0.5mm} 2} \epsilon \epsilon' - p_{z}' p_{z}).
\end{eqnarray}

\noindent The conservation of momentum relations $\epsilon' \vec{p}\hspace{0.5mm}' = \epsilon \vec{p} - \vec{k}$, and $\epsilon' \vec{\varepsilon}\hspace{0.5mm}' = \epsilon \vec{\varepsilon} - \vec{\omega}$ can be used to evaluate both contributions to the response tensor as follows. In the first term the primed quantities are replaced with unprimed quantities: $p' = \epsilon \epsilon' p - \epsilon' k$, $\varepsilon' = \epsilon \epsilon' \varepsilon - \epsilon' \omega$. In the second term the unprimed quantities are replaced with primed quantities: $p = \epsilon \epsilon' p' + \epsilon' k$, $\varepsilon = \epsilon \epsilon' \varepsilon' + \epsilon' \omega$. The result is:

\begin{eqnarray*}
\Pi^{12}_{\textrm{sd}} = -e^{2}  && \sum_{\epsilon, \epsilon'} \int  \frac{d^{3} \vec{p}}{(2\pi)^{3}} \frac{\epsilon \Delta n^{\epsilon}(\vec{p})}{\omega - \epsilon\varepsilon + \epsilon' \varepsilon' } \nonumber \\
&& \times \frac{im}{2\varepsilon \varepsilon' p}\left[\epsilon' (-\varepsilon^{2} + p^{2} +p_{z}^{2}) + \epsilon \epsilon' (\varepsilon \omega - p_{z} k_{z}) \right] \\
 +  e^{2} && \sum_{\epsilon, \epsilon'} \int  \frac{d^{3} \vec{p}\hspace{0.5mm}'}{(2\pi)^{3}} \frac{\epsilon' \Delta n^{\epsilon'}(\vec{p}\hspace{0.5mm}')}{\omega - \epsilon\varepsilon + \epsilon' \varepsilon' } \\
 && \times \frac{im}{\varepsilon \varepsilon' p'} \left[\epsilon (\varepsilon ^{2} - p'^{\hspace{0.5mm 2}} - p_{z}^{\prime \hspace{0.5mm 2}}) + \epsilon \epsilon' (\varepsilon' \omega - p_{z}' k_{z})\right]
\end{eqnarray*}

\noindent The next step is to evaluate the sums over $\epsilon' = \pm$ in the first term, and sums over $\epsilon = \pm$ in the second term. This requires the use of the following identities:
\begin{equation*}
\left[\frac{1}{\omega - \epsilon \varepsilon + \varepsilon'} - \frac{1}{\omega - \epsilon \varepsilon - \varepsilon'} \right] = \frac{-2\varepsilon'}{-2\epsilon pk + k^{2}},
\end{equation*}
\begin{equation*}
\left[\frac{1}{\omega - \varepsilon + \epsilon' \varepsilon'} - \frac{1}{\omega +  \varepsilon + \epsilon' \varepsilon'} \right] = \frac{2\varepsilon}{2\epsilon' p' k + k^{2}}.
\end{equation*}
The result is:
\begin{align}
&\Pi^{12}_{\textrm{sd}} =
\nonumber \\
&-ime^{2} \sum_{\epsilon} \int \frac{d^{3}\vec{p}}{(2\pi)^{3}} \frac{\Delta n^{\epsilon}(\vec{p})}{\varepsilon \varepsilon' p}  \times [p^{2} +p_{z}^{2} - \varepsilon^{2}] \left(\frac{-2\varepsilon'}{-2\epsilon pk + k^{2}} \right) \label{contrib1zero}\\
&-ime^{2} \sum_{\epsilon} \int \frac{d^{3}\vec{p}}{(2\pi)^{3}} \frac{\Delta n^{\epsilon}(\vec{p})}{\varepsilon \varepsilon' p}[\epsilon \varepsilon \omega - \epsilon p_{z}k_{z}] \left( \frac{-2\varepsilon'}{-2\epsilon pk + k^{2}} \right)\\
&+ime^{2} \sum_{\epsilon'} \int \frac{d^{3}\vec{p\hspace{0.5mm}'}}{(2\pi)^{3}} \frac{\Delta n^{\epsilon'}(\vec{p \hspace{0.5mm}'})}{\varepsilon \varepsilon' p'} [\varepsilon'^{2} - p'^{2} - p_{z}^{\prime \hspace{0.5mm} 2}] \left(\frac{2\varepsilon}{2\epsilon' p' k + k^{2}} \right) \label{contrib2zero}\\
&+ime^{2} \sum_{\epsilon'} \int \frac{d^{3}\vec{p\hspace{0.5mm}'}}{(2\pi)^{3}} \frac{\Delta n^{\epsilon'}(\vec{p \hspace{0.5mm}'})}{\varepsilon \varepsilon' p'}[\epsilon' \varepsilon' \omega - \epsilon' p_{z}' k_{z}]\left(\frac{2\varepsilon}{2\epsilon' p' k + k^{2}} \right).
\end{align}

\noindent By inspection of the identity in Eq. \ref{useinappendix1} it follows that $\varepsilon^{2} - p^{2} - p_{z}^{2} =0$ and $p^{2} + p_{z}^{2} - \varepsilon^{2}=0$. Note that the eigenvalues of the magnetic-dipole-moment spin operator are $sp$, where $s=\pm$. The next step is to swap the dummy variables present in the second contribution to the response tensor as follows: $p', \varepsilon', \epsilon' \rightarrow p, \varepsilon, \epsilon$. It is now clear that Eq. \ref{contrib1zero} and Eq. \ref{contrib2zero} do not contribute to the response, since they are equal to zero. The two contributions to the response tensor may now be combined into a single expression:
\begin{align*}
\Pi^{12}_{\textrm{sd}} =
 \; ime^{2} \sum_{\epsilon} \int & \frac{d^{3}\vec{p}}{(2\pi)^{3}} \frac{\epsilon \Delta n^{\epsilon}(\vec{p})}{\varepsilon p} [\varepsilon \omega - p_{z} k_{z}] \\ & \times \left(\frac{1}{-2\epsilon pk + k^{2}} - \frac{1}{2\epsilon pk + k^{2}} \right) \\
= \;  ime^{2}k^{2} \sum_{\epsilon}& \int  \frac{d^{3}\vec{p}}{(2\pi)^{3}} \frac{\epsilon \Delta n^{\epsilon}(\vec{p})}{(pk)^{2} - (k^{2}/2)^{2}} \left(\frac{-\varepsilon \omega + p_{z} k_{z}}{\varepsilon p} \right).
\end{align*}

By inspection of the above one recovers the 12-component appearing in Eq. \ref{spinmagresponse2}. The derivations of the other components of the response tensor follow in a similar way.

\nocite{*}

\bibliography{magbib}

\end{document}